\documentclass[iop,twocolumn, tighten, twocolappendix, numberedappendix]{emulateapj}

\usepackage[backref,breaklinks,colorlinks,citecolor=blue]{hyperref}        
\usepackage[all]{hypcap}

\usepackage{amsmath}
\usepackage{amssymb}
\usepackage{graphicx}
\usepackage{color}
\usepackage{natbib}
\usepackage{xspace}
\usepackage{nicefrac}
\usepackage[dvipsnames]{xcolor}
\newcommand\myshade{85}
\usepackage[T1]{fontenc}
\usepackage{lmodern}
 	
\colorlet{mycitecolor}{Turquoise}
\colorlet{mylinkcolor}{Turquoise}

\hypersetup{
  citecolor  = mycitecolor!\myshade!black,
  linkcolor  = mylinkcolor!\myshade!black,
  colorlinks = true,
}

\newcommand{\Msun}{M_\odot}

\newcommand{\Gyr}{\ensuremath{\,\mathrm{Gyr}}\xspace}

\newcommand{\Gpcyr}{\ensuremath{\, \mathrm{Gpc}^{-3}\mathrm{yr}^{-1}}\xspace}

\newcommand{\TheEvent}{The Event\xspace}

\def\Msun{{\rm M}_{\odot}}

\newcommand{\rr}[1]{#1}

\def\be{\begin{equation}}
\def\ee{\end{equation}}
\def\ba{\begin{eqnarray}}
\def\ea{\end{eqnarray}}

\defcitealias{Mandel+2016}{MdM16}

\submitted{Submitted}

\begin{document}

\title[Detectable Case M binary black hole mergers]{The chemically homogeneous evolutionary channel for binary black hole mergers: rates and properties of gravitational-wave events detectable by advanced LIGO}

\shorttitle{}

\author{ S. E. de Mink\altaffilmark{1} \& I. Mandel\altaffilmark{2} 
} 

 \affil{               $^{1}$Anton Pannekoek Institute for Astronomy, University of
Amsterdam, 1090 GE Amsterdam, The Netherlands (S.E.deMink@uva.nl)\\
     $^{2}$School of Physics and Astronomy, University of Birmingham, Birmingham, B15 2TT, United Kingdom
               (IMandel@star.sr.bham.ac.uk)\\ 
         }

\begin{abstract}
\noindent
We explore the predictions for detectable gravitational-wave signals from merging binary black holes formed through chemically homogeneous evolution in massive short-period stellar binaries.  We find that $\sim 500$ events per year could be detected with advanced ground-based detectors operating at full sensitivity.  We analyze the distribution of detectable events, and conclude that there is a very strong preference for detecting events with nearly equal components (mass ratio $>0.66$ at 90\% confidence in our default model) and high masses (total source-frame mass between $57$ and $103\, M_\odot$ at 90\% confidence).  We consider multiple alternative variations to analyze the sensitivity to uncertainties in the evolutionary physics and cosmological parameters, and conclude that while the rates are sensitive to assumed variations, the mass distributions are robust predictions.  Finally, we consider the recently reported results of the analysis of the first 16 double-coincident days of the O1 LIGO (Laser Interferometer Gravitational-wave Observatory) observing run, and find that this formation channel is fully consistent with the inferred parameters of the GW150914 binary black hole detection and the inferred merger rate.

\end{abstract}

\section{Introduction}

The detection of the gravitational-wave signal GW150914 on September 14, 2015 from the inspiral and merger of two black holes with masses around $30\Msun$ by the Laser Interferometer Gravitational-wave Observatory (LIGO) has provided the first robust evidence that black holes with such masses exist, that they can form in binary pairs, and that they coalesce at an inferred local rate of 2--400\Gpcyr \citep{LIGO+2016_mainPaper, LIGO+2016_astroImplications,LIGO+2016_rate}.   

Predictions for the rate of binary black hole mergers varied widely due to the lack of direct observational evidence \citep{Abadie+2010}.  Empirical estimates are available for the merger rates of binary neutron stars, based on the observed populations of double neutron stars \citep[e.g.][]{Phinney:1991ei,Narayan:1991,Kim:2003kkl,OShaughnessyKim:2009}.  In contrast, for double black hole mergers the population of direct progenitors is not accessible and the rate prediction fully relied on the predictions of stellar and binary evolutionary models integrated into population synthesis simulations.  Several groups predicted that the gravitational-wave signals of binary black hole mergers would potentially dominate LIGO observations \citep[e.g.][]{Lipunov+1997,Voss+2003, Belczynski+2010a, Dominik+2015}, but these analyses also demonstrated the significant uncertainties in these predictions \citep[e.g.][]{Dominik+2012,de-Mink+2015}.  

The detection of GW150914 within the first 16 days of the advanced LIGO O1 observing run with two detectors operating in coincidence has provided the first stringent empirical constraints of the binary black hole merger rate.  Assuming the results are representative, this implies the possibility of hundreds of detections per year as the detectors reach full design sensitivity and the duration of the runs with both detectors online increases \citep{scenarios, LIGO+2016_rate,LIGO+2016_astroImplications,LIGO2010_DOCanticipatedSensitivity}.  This will make it possible to constrain the formation channels for what is by far the most intriguing outcome of massive binary evolution, the coalescence of two gravitational singularities \citep[e.g.,][]{BulikBelczynski:2003,MandelOShaughnessy:2010,Stevenson:2015,Mandel+2015}.

Different channels have been proposed for the formation of double black hole binaries that can coalesce within a Hubble time.  These include: 
\begin{enumerate}
\item[(i)] \emph{dynamical formation}, which requires a dense star cluster \citep[e.g.,][]{Sigurdsson+1993, Portegies-Zwart+2000,MillerLauburg:2008,Rodriguez:2015, Antonini+2016}.
\item[(ii)] \emph{classical isolated binary evolution} through highly non-conservative mass transfer or common envelope ejection \citep[e.g.,][]{Tutukov:1973, Tutukov:1993, Kalogera+2007, Belczynski+2016}; and 
\item[(iii)] \emph{chemically homogeneous evolution in tidally distorted binary stars}, i.e.~massive stars in (near) contact binaries that experience strong internal mixing as proposed by \citet{de-Mink+2008, de-Mink+2009a} and further explored in the context of the formation of binary black holes by \citet{Mandel+2016} and \citet{Marchant:2016}.  
\end{enumerate}

The third channel, the topic of this paper, originates from very close binary systems that are in (near) contact at the onset of hydrogen burning.  In such systems, the deformation by tides of the component stars triggers instabilities in the stellar interior that can, in principle, drive large-scale Eddington-Sweet circulations  \citep{Endal+1978, Zahn1992}. This allows mixing of nuclear burning products produced in the center throughout the stellar envelope. \rr{Originally these processes have been considered in the case of rotating single stars and have been proposed as an explanation for surface abundance anomalies such as nitrogen enhancements (e.g., \citealt{Maeder+2000a}, see however \citealt{Brott+2011a}).}  

If the large-scale circulations are efficient enough, they will lead to a gradual enrichment of the stellar envelope with helium. This prevents the buildup of a chemical gradient between core and envelope that characterizes non rotating stars in the classical evolutionary models. This mode of evolution is referred to as ``chemically homogeneous evolution'', originally proposed for rotating single stars by  \citet{Maeder1987}. The stars are well approximated by the classical homology relations, i.e., the approximate analytic scaling solutions for the stellar structure equations which assume a uniform chemical composition.  They stay compact during their main sequence evolution as they slowly evolve towards the helium main sequence.  This mode of evolution gained renewed attention in the context of the formation of the progenitors of long gamma-ray bursts from rapidly rotating single stars \citep{Yoon+2005, Woosley+2006}.  \rr{Solid evidence is missing, but observations have provided several clues in favor of the existence of chemically homogeneously evolving stars, based on individual objects  \citep{Martins+2013,Almeida+2015} as well as the properties of unresolved populations \citep{Eldridge+2012, Stanway+2014, Szecsi+2015}, as discussed by \citet[][see Section 2.4]{Mandel+2016}. }
 
Here we discuss the evolutionary channel proposed by \citet{de-Mink+2009a}, who argued that the conditions for chemically homogeneous evolution can, in principle, be achieved in very close massive binary systems. The classical evolutionary models predict that near contact binaries with orbital periods less than about 2 days will merge even before or soon after the completion of hydrogen burning due to the expansion of the stars \citep{Nelson+2001,de-Mink+2007}. On the other hand, models that account for enhanced mixing allow for the possibility that the two stars shrink and remain within their Roche lobes.  This evolutionary channel has been explored with three different 1D evolutionary codes \citep{de-Mink+2009a,Song:2016, Marchant:2016}. All three studies report the existence of a window in the initial binary parameter space for this type of evolution when accounting for mixing induced by rotation and angular momentum transport by magnetic fields. The latter group even explores the evolution of over contact systems. Examples of observed binary systems that have been proposed to undergo (partial) chemically homogeneous evolution are VFTS 352 \citep{Almeida+2015} and HD 5980 \citep{Koenigsberger+2014}.

This channel naturally produces rather massive binary black holes as the stars process a larger fraction of their initial mass by nuclear fusion.  The allowed initial binary parameter space further favors producing binary black holes with comparable masses. The black holes thus formed already reside in a close orbit, so that most of them coalesce within a Hubble time as the orbit decays due to gravitational wave radiation. The limiting factor comes from the stellar wind mass loss, which affects the final masses as well as the final orbital separation, and can potentially inhibit chemically homogeneous evolution if the binary expands to the point that the stars significantly spin down. The reduction of stellar wind mass loss at low metallicity \citep{Vink+2001,Mokiem+2007a} leads to a preference for the progenitors to form at higher redshift or in dwarf galaxies. 

The Monte Carlo simulations by \citet{Mandel+2016} of the cosmological merger rate through this channel imply delay times of 3--11Gyr, a preference for comparable mass ratios $q>0.75$ and typical total masses near 50--110$\Msun$ in the default model considered there. These simulations predict a local $z=0$ merger rate of 10\Gpcyr, peaking at a redshift of 0.5 at twice the local rate, implying that the channel can be potentially the dominating channel for binary black hole in this mass regime.  The detailed 1D evolutionary models by \citet{Marchant:2016} \rr{account for} over contact systems, which produce mass ratios closer to unity, higher total masses, a larger range of delay times and somewhat lower rates due to the stronger preference for low metallicity. 

The aim of this paper, which is a companion paper to \citet{Mandel+2016}, is to provide the expected detection rates as well as the distributions of masses, mass ratios and chirp masses of detectable events that form through this channel. We provide estimates for the anticipated final design sensitivity, as well as for the lower sensitivity achieved during the 16 day portion of the O1 run which led to the detection of ``The Event'' GW150914.  We compare the parameters inferred for GW150914 with the estimates and conclude they are fully consistent. We discuss the impact of variations of the model assumptions and show that, even though the rates are substantially uncertain, the preference for high masses and mass ratios similar to GW150914 is a robust prediction of this channel.  

We make the full output of our simulations available online for the community at \url{http://www.sr.bham.ac.uk/~imandel/CaseM}, in order to allow further comparisons with current and future data and with simulations of other channels.

\section{Model Assumptions} \label{sec:inputs}

Our simulations of massive close binary populations over cosmic time are described in \citet[][hereafter \citetalias{Mandel+2016}]{Mandel+2016}, to which we refer for a full description. Here, we summarize the key assumptions.

\subsection {\rr{Progenitor evolution}}
We perform a Monte Carlo simulation drawing the initial parameters of massive binary systems from a Kroupa initial mass function (IMF) for the primary star \citep{Kroupa+2003}, a flat mass ratio distribution \citep[e.g.][]{Sana+2012, Kobulnicky+2014} and a distribution of orbital periods appropriate for O-type stars \citep{Sana+2012} as detailed in \citetalias{Mandel+2016}. We follow the evolution of the systems, parametrizing our assumptions as described below.  We check if the stars fit  within their Roche lobes at zero age using the radii of zero age main sequence stars based on models computed with Eggleton's evolutionary code \citep{Pols+1995, Glebbeek+2008}.  We assume that the stellar spin is synchronized with the orbit and the orbits are circular, which is appropriate for the short period systems of interest \citep{Zahn1989}. Using the spin frequency and stellar radius we compute the fraction of the Keplerian rotation rate, which we compare with the threshold for chemically homogeneous evolution in the detailed models by \citet{Yoon+2006}. These are 1D hydrodynamical evolutionary models that solve the stellar structure equations accounting for the effect of the centrifugal acceleration and rotationally driven instabilities \citep{Endal+1976,Heger+2000}, which lead to the transport of chemical elements and angular momentum. These models further account for internal magnetic fields \citep{Spruit2002}.  For the threshold for chemically homogeneous evolution we use the expression given in Section 4.3 of \citetalias{Mandel+2016}. Following \citet{Yoon+2006}, we adopt a maximum metallicity threshold of $Z=0.004$.  
%XXX

If the system fulfills the criteria for chemically homogeneous evolution, we follow the evolution by accounting for the effects of mass and angular momentum loss driven by stellar winds and the final supernova explosions via the simple parametrized approach described in \citetalias{Mandel+2016}.  We account for the effect of mass loss on the orbital separation and the masses of the final remnants.  We only consider systems in which both stars fulfill the threshold for chemically homogeneous evolution throughout their main-sequence evolution. We consider the possibility that the most massive helium stars lead to pair-instability supernovae leaving no remnant, by adopting an upper limit of 63$\Msun$ \citep{Heger+2002} for the final, pre-explosion mass of the star.  \rr{We do not consider systems that are more massive than the pair instability regime, in contrast to \citet[][]{Marchant:2016}, given the lack of constraints on the progenitors of systems in this mass range. We further conservatively exclude any additional contribution from systems that evolve through an over-contact phase.}   

Given the high orbital velocities in the massive close systems \rr{under consideration, we ignore} the effect of possible natal kicks accompanying collapse to black holes.  We account for the decay of the orbit by energy and angular momentum loss due to the emission of gravitational waves as in \citet{Peters1964}. 

\rr{Our approach differs from the complementary work of \citet{Marchant:2016}, who explicitly follow the full evolution with a stellar evolutionary code.  However, given the large uncertainties in evolutionary models and the mixing processes in particular, we have opted for a faster parametrized approach which allows us to study the effect of various uncertainties in \autoref{variations}.}

\subsection{Cosmology}\label{cosmo}

To compute the cosmological merger rate history we adopt a  standard flat cosmology with $\Omega_\Lambda = 0.718$ and $h_0=0.697$ \citep{WMAP9}.  We adopt the star formation rate $d^2 M_{\rm SFR} / (dt  dV_{\rm c}) \, (z)$ per unit source time per unit comoving volume as a function of redshift $z$ from \citet[][Eq.~15 in their work]{MadauDickinson:2014}.  For the metallicity distribution as a function of redshift, we follow \citet{LangerNorman:2006}, which is based on the mass--metallicity relation of \citet{Savaglio:2005} and the average cosmic metallicity scaling of \cite{KewleyKobulnicky:2005,KewleyKobulnicky:2007}.   For the average present day metallicity we conservatively use $1.06$ times solar metallicity, $Z_\odot=0.0134$ \citep{Asplund:2009}. We implicitly assume that the initial mass function and other binary properties do not depend on metallicity or redshift. This is reasonable since the fraction of binaries of interest formed in metal-free population III stars or extremely metal-poor  stars is very small within this framework of assumptions and the merger rate is dominated by systems with metallicity near the maximum threshold metallicity.  For this metallicity observations indicate no evidence for a varying IMF \citep{Kroupa2002}. 

The rate density of binary black hole mergers is given in \citetalias[][Eq.~(8),]{Mandel+2016} as the number of mergers $N_{\rm merge}$ per unit component mass $m_1$ and $m_2$ at the moment of merger $t_{\rm m}$ per unit source time and per unit comoving volume $V_c$

\begin{eqnarray}
\nonumber
\frac{d^4 N_{\rm merge}}{dV_{\rm c} \, dt \, dm_1 \, dm_2} (t_{\rm m})= \int_{P_{\min}}^{P_{\max}} dP  \int dZ  \int_0^{t_{\rm m}} dt \ \ \Big\{\\
\nonumber
 p(t_{\rm m}; m_1, m_2, P, Z, t_{\rm b}) \,\, \frac{d^2M_{\rm SFR}}{dt \, dV_{\rm c}} (t_{\rm b})  \\
 \times \frac{d^5 N_\textrm{binaries}}{dm_1 \, dm_2 \, dP \, dZ \, dM_{\rm SFR}} (t_{\rm b}) \Big\} \, .
 \end{eqnarray}
Here, the star formation rate $ {d^2M_{\rm SFR}}/(dt \, dV_{\rm c})$ is evaluated at the binary birth time $t_{\rm b}$ and  ${d^5 N_\textrm{binaries}}/({dm_1 \, dm_2 \, dP \, dZ \, dM_{\rm SFR}})$ is the number density of binaries formed per unit $m_1$, $m_2$, initial orbital period $P$, and metallicity $Z$ per unit star formation rate.  The time delay distribution is given by the probability density $p(t_{\rm m}; m_1, m_2, P, Z, t_{\rm b})$ for a binary to merge at time $t_{\rm m}$ if it was formed with the given $m_1$, $m_2$, $P$, $Z$ at time $t_{\rm b}$.  Note that  $m_1$ and $m_2$ refer to the black hole masses and not the birth masses of the progenitor stars.  The innermost integral is taken over all birth times  $t_{\rm b}$ preceding the merger time $t_{\rm m}$, where the zero of time corresponds to the Big Bang.  \rr{The minimum and maximum initial orbital periods $P_{\min} = 10^{0.075}$ days and $P_{\max} = 10^{5.5}$ days and the initial period distributions are based on observations of O-type stars \citep{Sana+2012}, extending the period distribution to allow for effectively single stars as in \citet{de-Mink+2015}.} For further information we refer to \citetalias{Mandel+2016}.

\begin{figure}[bt]\center
  \includegraphics[width=0.5\textwidth]{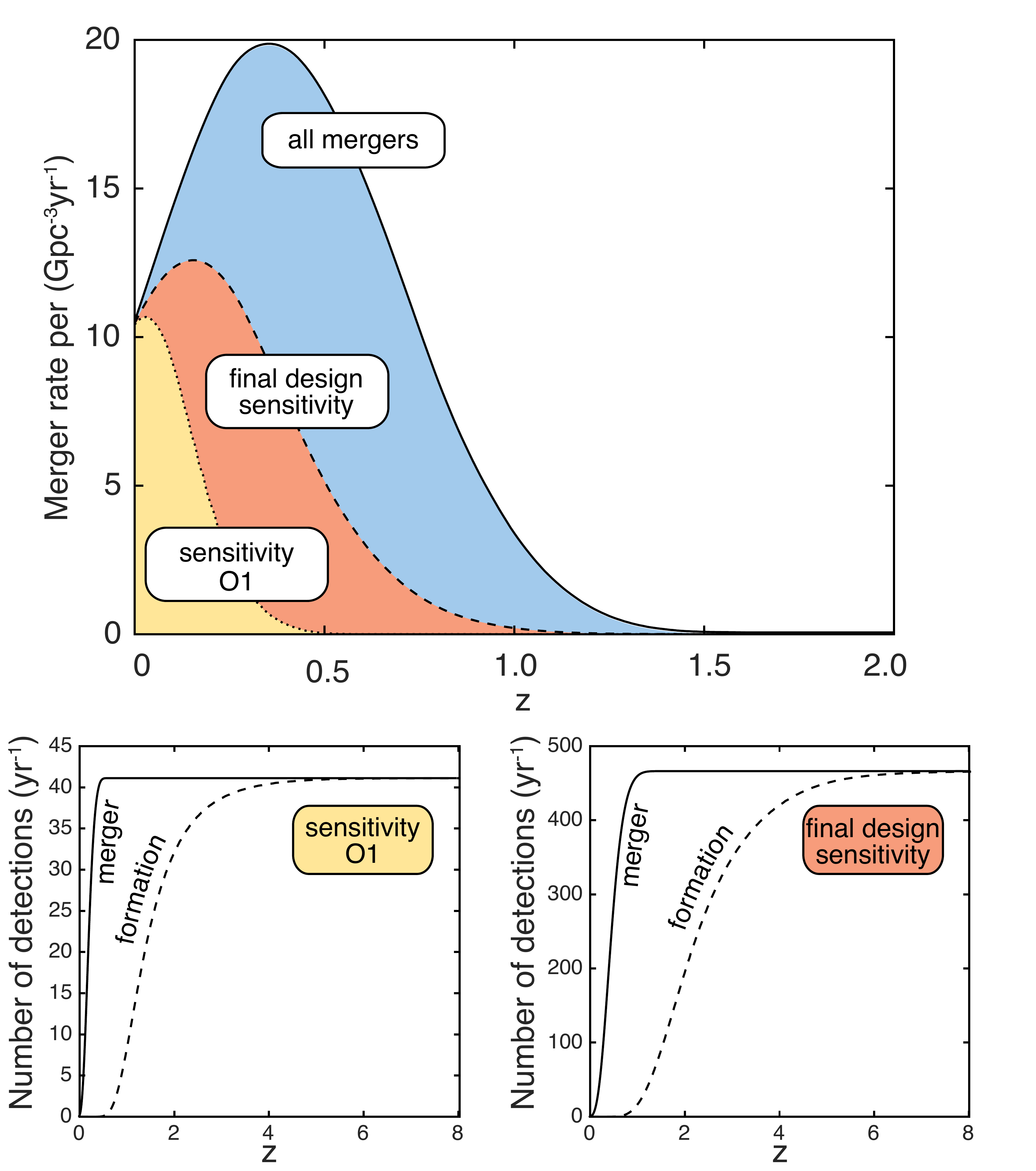}
  \caption{ Merger rate as a function of redshift for binary black hole mergers formed through the chemically homogeneous evolutionary channel (solid line, blue shaded) together with the merger rate of detectable events for the final design sensitivity (dashed line, orange shading) and O1 sensitivity  (dotted line, yellow shading). The rate is given in \Gpcyr in the comoving frame.  In the lower panels we provide the cumulative distribution of the total number of detections as a function of merger redshift (solid line) as well as the cumulative distribution of detectable systems by the redshift of formation of stars in the binary system that gave rise to the merger (dashed line).    %
    \label{CosmicMergerRate}}
\end{figure}

\subsection{Detection rates}

We convert the cosmological merger rate into a rate of detections by advanced LIGO \citep{AdvLIGO} and Virgo \citep{AdvVirgo} detectors per unit observer time by folding in the gravitational waveform models and the detector sensitivity:

\begin{eqnarray}
\nonumber
R_{\rm detect} \equiv \frac{dN_\textrm{detect}}{dt_{\rm obs}} =   \int_0^{\infty} dz \int dm_1  \int dm_2 \, \Big\{  \\
\nonumber
 f_{\rm detect}(z(t_{\rm m}),m_1,m_2) \frac{dV_{\rm c}}{dz}   \frac{1}{1+z} \times
\\
\frac{d^4 N_{\rm merge}}{dV_{\rm c} \, dt \, dm_1 \, dm_2} \,  (t_{\rm m}(z)) \,  \Big\} \, .
\end{eqnarray}
\vspace{0.5mm} 

\noindent Here, $ f_\textrm{detect} = f(z(t_{\rm m}),m_1,m_2) $ is the probability that LIGO and Virgo will detect a coalescing black hole binary with given component masses $m_1$ and $m_2$ merging at a redshift $z(t_{\rm m})$ such that the gravitational waves emitted from a source merging at time $t_m$ will arrive at the Earth today. The term  $1/(1+z)$ reflects the time dilation of the source clock (with which all times are measured unless otherwise specified) with respect to the observer clock $t_{\rm obs}$.  We evaluate these integrals with a Monte Carlo simulation.

We model the gravitational-wave emission from a binary by using the {\tt IMRPhenomB} waveform approximant \citep{Ajith:2011}.  This waveform includes the post-Newtonian inspiral and the perturbative ring down, connected by a smooth merger via a phenomenological approximation calibrated to numerical relativity simulations.  Although we expect the spins of the stars to be aligned by tides for this formation channel, the orientation and magnitude of the spins of the black holes are uncertain as they may be affected by stochastic processes during the collapse. For the purpose of the waveform calculation we set the spins to zero. This generally underestimates the strength of the gravitational-wave signal relative to that expected from a binary with aligned spins, which seems likely in this scenario \citepalias{Mandel+2016}.  Although more precise waveforms are now available, the accuracy provided by {\tt IMRPhenomB} is sufficient for our purposes here, especially since most of our binaries have mass ratios close to unity.  

As described in \citepalias{Mandel+2016}, our Monte Carlo simulation generates a set of simulated merging binary black holes, which we label with an index $k$.  We divide the history of the Universe into a large number of bins by redshift, which we label with an index $j$.   For each sample binary $k$, we redshift the waveform to account for the cosmological expansion of the Universe, thus producing a redshifted frequency-domain waveform $\tilde{h}(f)_{k,j}$ for each merger bin $j$ corresponding to redshift $z_j$.  We compute the signal-to-noise ratio $\rho$ at which an optimal (face-on, overhead) source at this redshift and its corresponding luminosity distance $d_L(z_j)$ would be detected by a single advanced LIGO interferometer:
\be
\rho_{k,j}^2 = 4 \int_0^\infty \frac{|\tilde{h}(f)_{k,j}|^2}{S_n(f)} df\, .
\ee
Here, $S_n(f)$ is the noise power spectral density of the detectors.  To estimate detectability at full design sensitivity, we use the so-called zero-detuning, high-power configuration \citep{LIGO2010_DOCanticipatedSensitivity}.  For estimates at O1 sensitivity, we use the reference O1 noise curve \citet{LIGO2015_DOCcalibratedO1sensitivity} (see \citet{LIGO+2016_calibration} for associated calibration accuracy).

The signal-to-noise ratio will depend on the source location on the sky relative to the detector and the source orientation. The projection coefficient $\Theta$ as a function of these angles is given by \citet{Finn:1996}.  We choose a single-detector threshold signal-to-noise ratio $\rho_{t} = 8$  as a proxy for the detectability of the source by a network; the detection probability for the given source at a given redshift is then \citep[e.g.,][]{Belczynski:2014VMS}
\be
f^{\rm detect}_{k,j} = 1- \mathcal{C}_{(\Theta/4)}\left[\min\left(\frac{8}{\rho_{k,j}}, 1\right)\right],
\ee
where the cumulative distribution function of the projection coefficient, $\mathcal{C}_{(\Theta/4)}$, is measured with a separate numerical Monte Carlo.

We can finally compute the total merger rate that is detectable by summing over all redshifts bins and simulated binaries:
\be
R_{\rm detect} = \sum_k \sum_j f^{\rm detect}_{k,j} \, \frac{dN_{k,j}^{\rm merge}}{dt \, dV_{\rm c}} \,  dV_{\rm c} (z_j) \, \frac{1}{1+z_j},
\ee
\rr{where $dN_{k,j}^{\rm merge}/ (dt \, dV_{\rm c})$ is the merger rate for sample binary $k$ in redshift bin $z_j$}, $dV_{\rm c}(z_j)$ is the comoving volume associated with redshift bin $z_j$ and the last term comes from the difference between source time and observer time, $dt / dt_{\rm obs} = 1/(1+z)$.

\begin{figure*}[t]\center
      \includegraphics[width=\textwidth]{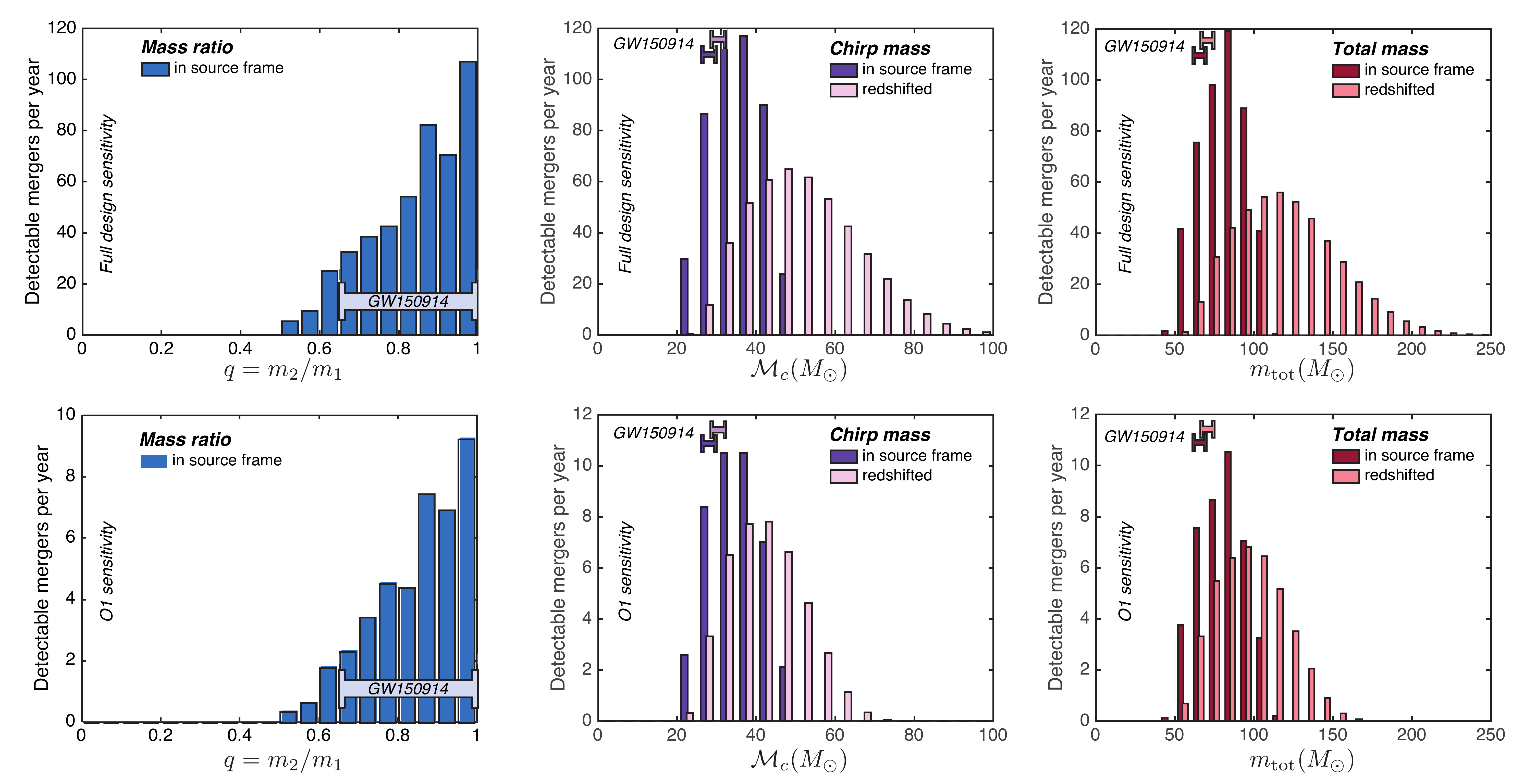}%
    \caption{ Distributions of the parameters that characterize the masses of the two black holes  that can be inferred for detectable events for full design sensitivity (top row) and O1 sensitivity (bottom row).  Distributions of the mass ratio $q = m_2/m_1$, chirp mass $\mathcal{M}_c = m_1^{3/5} m_2^{3/5} (m_1+m_2)^{-1/5}$ and total mass $m_1 + m_2$ are given, together with the redshifted values in the last two cases.   Approximate uncertainty intervals corresponding to the inferred parameters of \TheEvent are indicated with error bars \citep{LIGO+2016_properties}.  \label{1D} }
\end{figure*}

\section{Results}\label{sec:results}

We provide predictions for the number of detectable events and their distribution, and compare these with observed values based on \TheEvent, GW150914.  These results are summarized in \autoref{tab1}.

\subsection{Cosmic merger rate and the local merger rate} 
The overall cosmic merger rate is shown in \autoref{CosmicMergerRate} for our default simulation.  The shape follows the rise and fall of the cosmic star formation rate for low metallicity stars \citepalias[see also Fig.~3 \& 8 in][]{Mandel+2016}, shifted by the time delay between the the birth of a massive binary star progenitor system and the final merger of the two black holes.  The typical time delay in our default model is  4-11\Gyr \citepalias[cf., Fig.~6 in][]{Mandel+2016}.   As a result, the earliest mergers occur at a redshift of $z \sim1.5$, and the merger rate reaches a maximum of about 20\Gpcyr at $z \sim 0.4$ after which it drops by a factor 2 at $z = 0$. 

The local merger rate derived from our default model, 10\Gpcyr and the estimates obtained when we vary our model assumptions (see Table 1 of \citetalias{Mandel+2016}) are consistent \footnote{With the exception of a model variation {\tt Mdot2}, which produces zero detectable events from this channel as we discuss below.}  with the conservative inferred range of $2$--$400\Gpcyr$ from 16 days of double coincident advanced LIGO O1 observations \citep{LIGO+2016_rate}. The inference is based on \TheEvent as well as lower-significance triggers assuming a redshift-independent volumetric merger rate. The ranges allow for different underlying mass distributions for the BH-BH population. 

\subsection{Detection rate} 

Our default model predicts that advanced gravitational-wave detectors operating at full design sensitivity could observe $470 \pm 25$ events per year of coincident observation resulting from binary black hole mergers formed through the Case M scenario. The error bar given here corresponds exclusively to the numerical uncertainty of the Monte Carlo integral, and does not include the systematic uncertainties in the assumed model, which are discussed in the next section. The corresponding rate for the sensitivity of the first observing run implies about 40 events per year. This scales to 1--2 detections for the first 16 days of double-coincident O1 observations. 

The 16 day double-coincident O1 run yielded one significant detection as well as one candidate event of lower significance, which has a posterior probability larger than 0.8 to be of astrophysical origin. No other triggers with significance larger than 0.5 were reported \citep{LIGO+2016_rate}.  These findings are consistent with the prediction of 1-2 events in our default model and the ranges obtained when exploring variations discussed \autoref{variations}.

\subsection {Redshift distribution of detectable mergers}

The reach of gravitational-wave instruments is limited. The gravitational-wave strain and, hence, the signal-to-noise ratio are inversely proportional to the luminosity distance \rr{at fixed redshifted masses $m (1+z)$ (see below)}.  Therefore, detection efficiency drops as a function of distance (redshift), with only massive and favorably located and oriented sources detectable at higher redshifts \rr{\citep[see Figure 4 of][]{LIGO+2016_astroImplications}}.  As a result, the redshift distribution of detectable events is shifted toward lower redshifts with respect to the total merging binary population. This is shown in  \autoref{CosmicMergerRate}.  The corresponding cumulative distributions of the redshift of detectable events are shown in the lower panels of  \autoref{CosmicMergerRate}. 

The median redshift for detectable sources is $z\sim0.5$ in our default simulation for full design sensitivity. During the less sensitive O1 run we are biased to events occurring at smaller redshifts and the median redshift of detections is $z \sim 0.2$.   The redshift inferred for \TheEvent, $z = 0.09^{+0.03}_{-0.04}$ \citep{LIGO+2016_properties}, lies approximately at the lower tenth percentile of the simulated distribution of detectable mergers for O1 sensitivity. 

We also provide the cumulative distribution of the redshift of formation for the detectable events in the lower panels of  \autoref{CosmicMergerRate}.  The typical events observable at full sensitivity result from systems that were formed at redshifts $z\sim$1--4.8 (90\% range), implying that they probe star formation and massive star evolution during and prior to the cosmic star formation peak.  

\citetalias{Mandel+2016} found that a total of $\sim 1250$ binary black holes merge per year of local ($z = 0$) observer time after forming through the chemically homogeneous evolution channel.  The detection rate calculations described above indicate that $\sim 40\%$ ($\sim 3\%$) of all potentially observable mergers could be detected with instruments operating at full design (O1) sensitivity.

\begin{figure*}[bt]\center
  \includegraphics[width=0.4\textwidth]{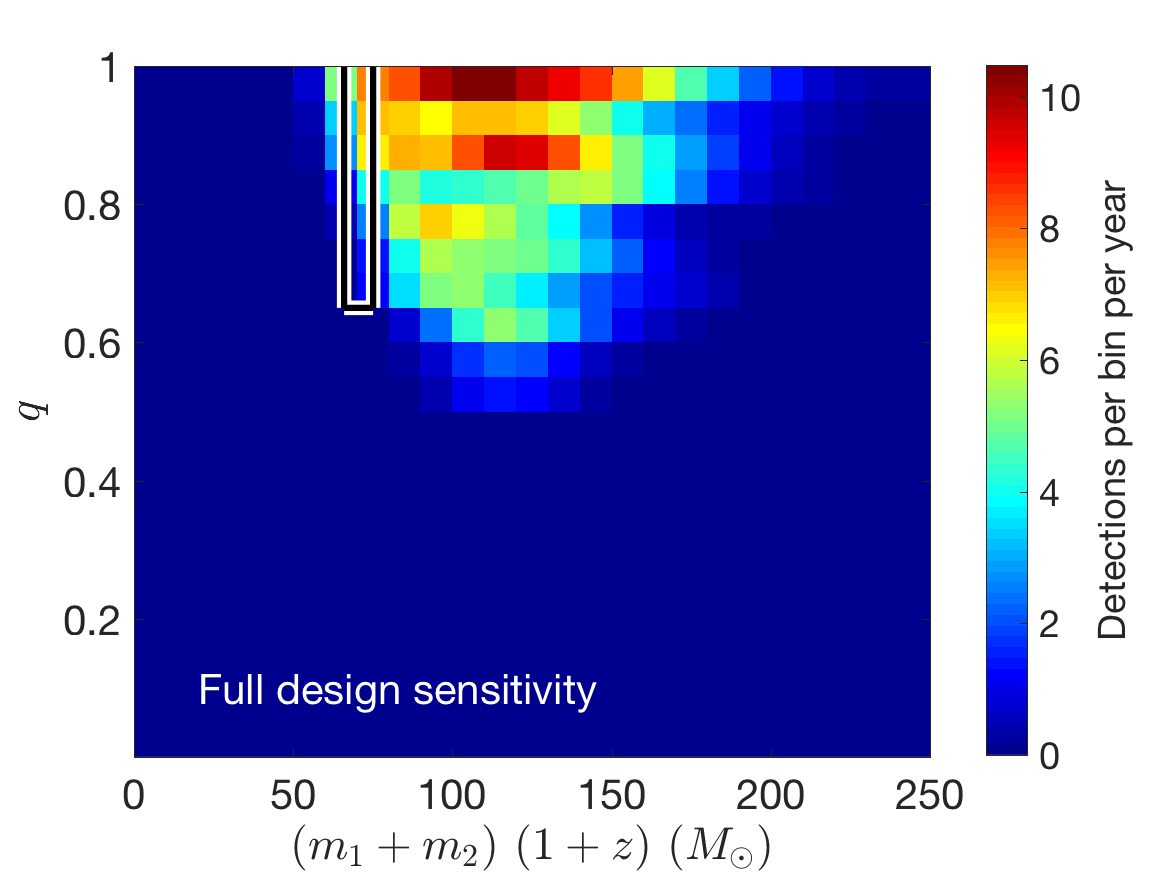}%
  \includegraphics[width=0.4\textwidth]{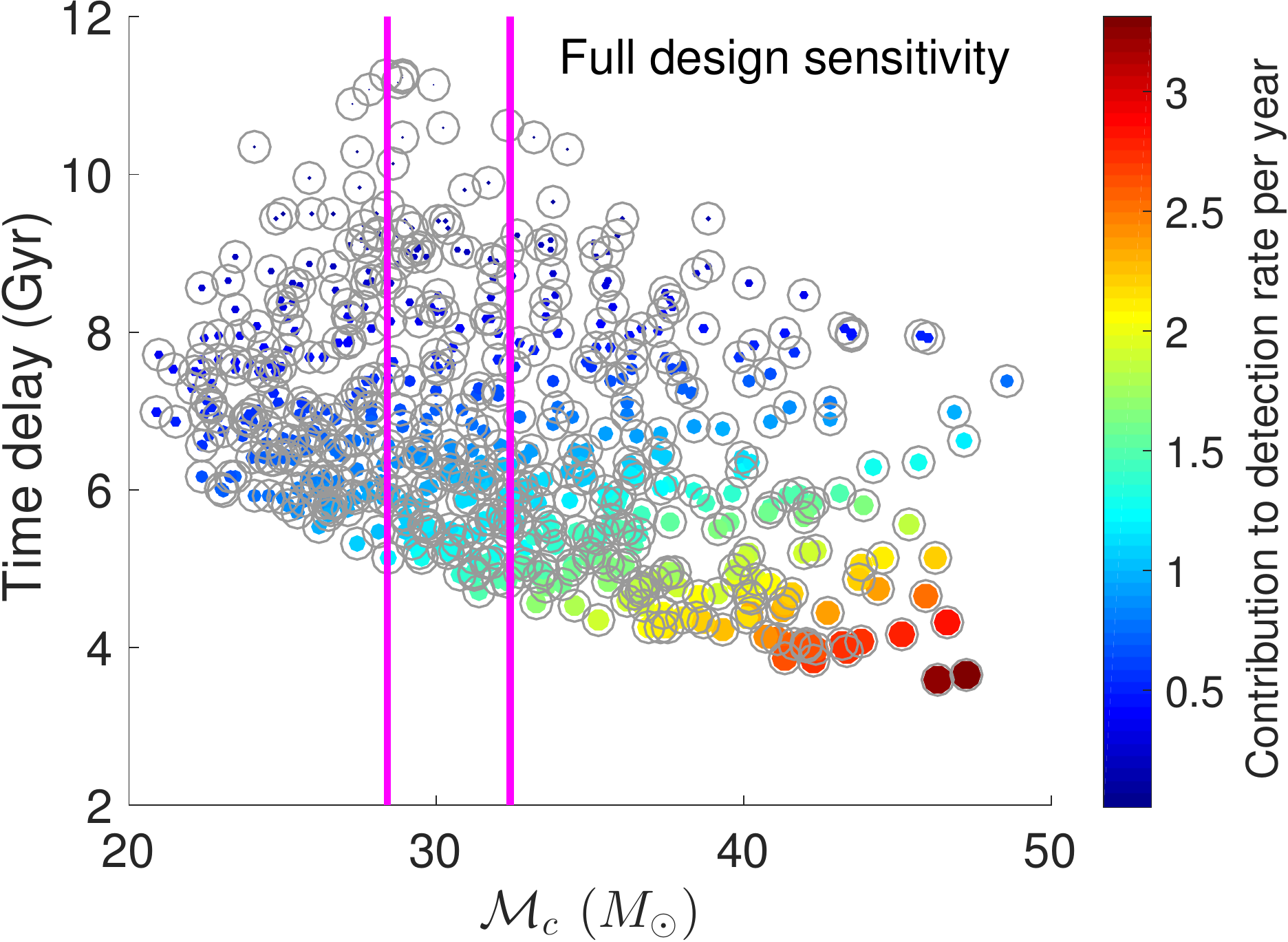}
  \includegraphics[width=0.4\textwidth]{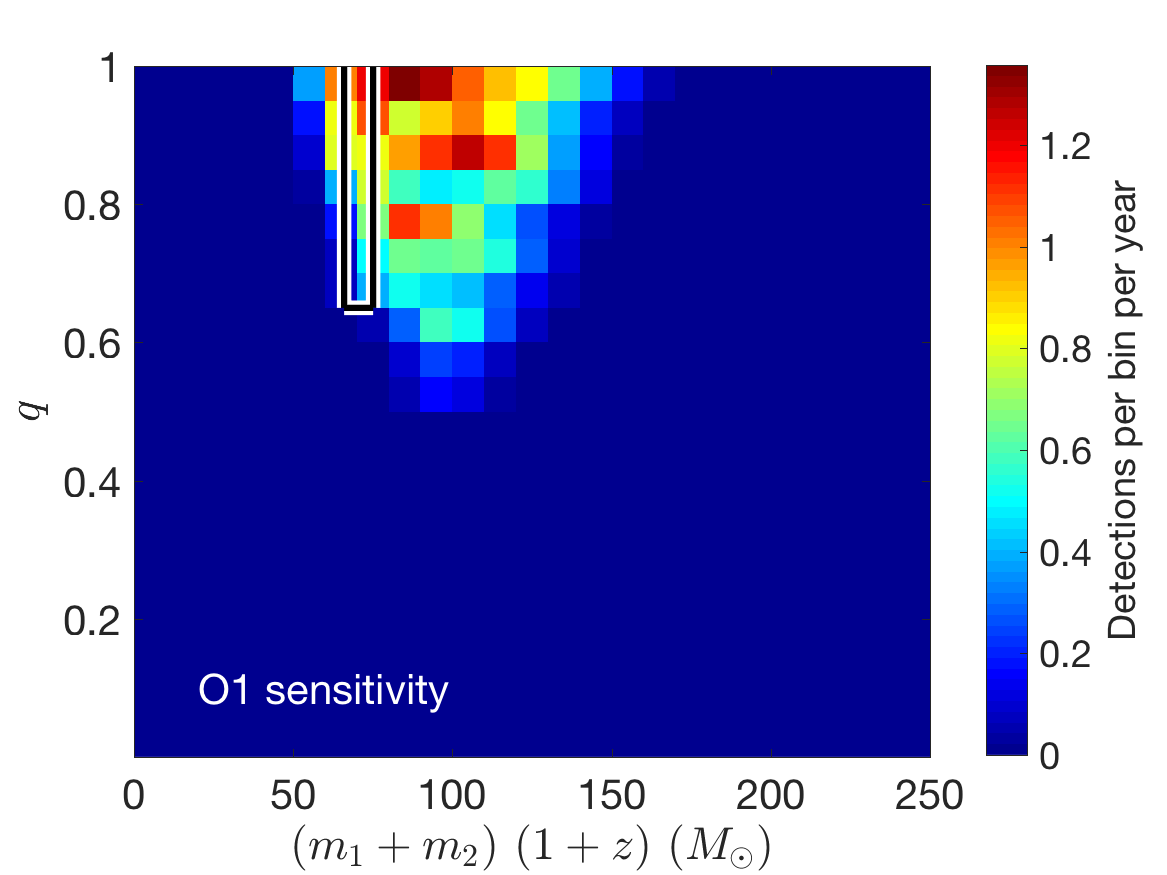}%
\includegraphics[width=0.4\textwidth]{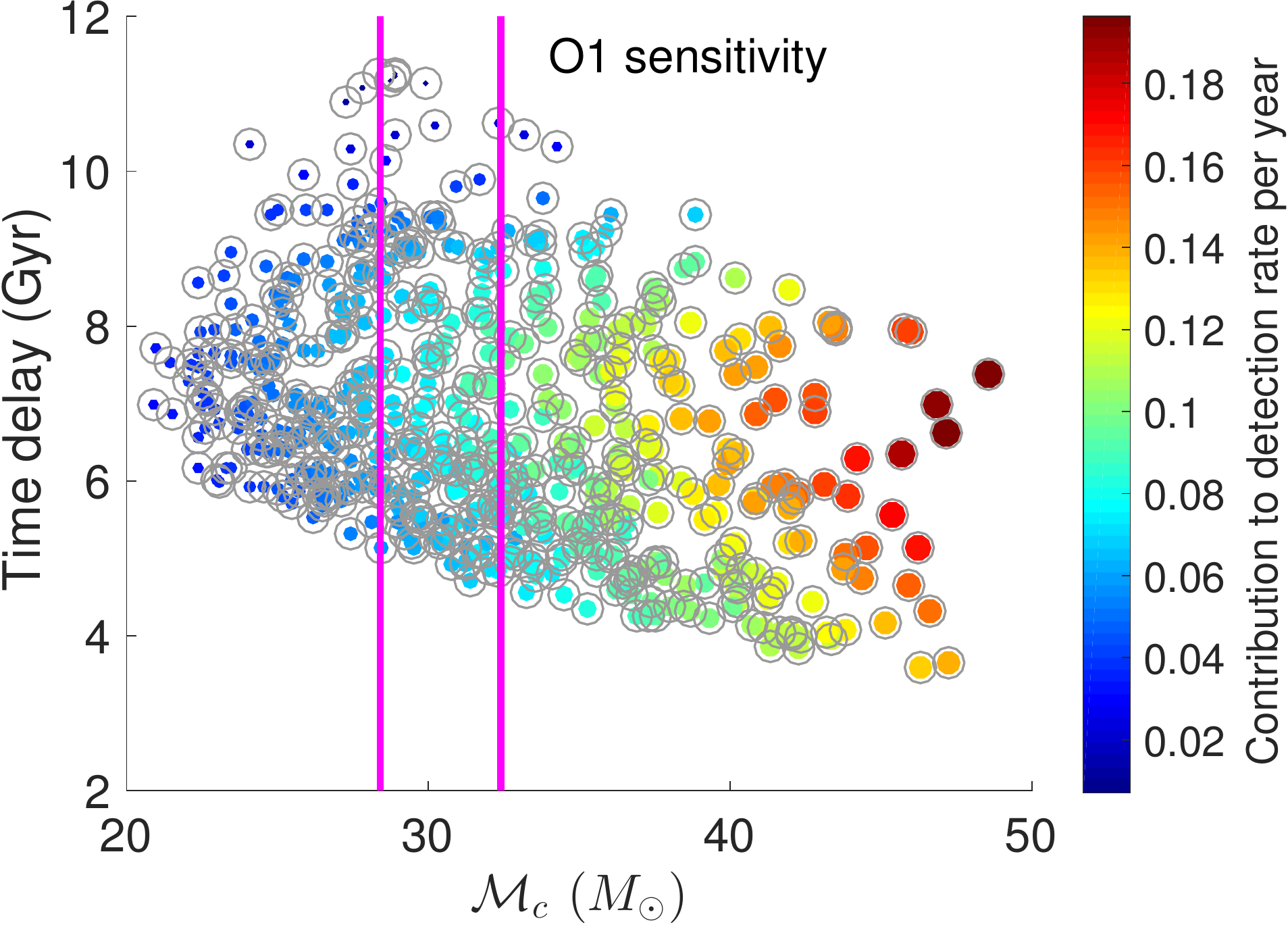}
  \caption{Two-dimensional parameter distributions are shown for detectable systems at full design sensitivity (top panels) and O1 sensitivity (bottom panels).   The left-hand panels shows the joint 2D distribution of the mass ratio $q = m_2/m_2$ and redshifted total mass $(m_1 + m_2) (1+z)$. Color shading indicates the number of detectable events per bin of 0.05 in q and $10\Msun$ in mass.  
The right-hand panel shows the time delay and chirp masses of individual simulated binaries in our Monte Carlo simulation. The color and size of the symbol indicate the contribution of these systems to the detection rate.  The approximate uncertainty intervals corresponding to the inferred \TheEvent parameters are overplotted in both panels.
\label{Correlation}}
\end{figure*}

\subsection{Total masses, chirp masses and mass ratios}
In \autoref{1D} we show the predicted distributions of properties that can in principle be inferred from the gravitational-wave signals of detected sources. Distributions for sources detectable at full design sensitivity are shown in the top row and the predictions for those detectable at O1 sensitivity are in the lower panels, together with the inferred parameters of GW150914.  

There is a strong preference for events resulting from systems with comparable masses for the individual black holes. 
There are no binaries of interest with mass ratios $q = m_2/m_1<0.5$ and more than two thirds of detections come from sources with $q>0.8$, as can be seen in the left-hand panels in \autoref{1D}. The preference for equal masses is a robust prediction of this evolutionary scenario  (\autoref{variations}) and is independent of the assumed detector sensitivity. The inferred mass ratio for \TheEvent is consistent with these predictions.

We further show the distributions for the chirp mass, $\mathcal{M}_c = m_1^{3/5} m_2^{3/5} (m_1+m_2)^{-1/5}$, and total mass, $m_{\rm tot} = m_1 + m_2$  in the central and right-hand panels of  \autoref{1D}.  The chirp mass is a combination of component masses $m_{1,2}$ which governs the phase evolution of gravitational waves at the leading order during the inspiral phase, and is therefore the most readily observable parameter for low-mass binaries.   However, for high-mass systems of interest here, typically only the late stages of the inspiral fall within the sensitive frequency band of the detectors. The total mass therefore becomes the more accurately measurable mass parameter \citep{Veitch:2015,Graff:2015,Haster:2015IMRI}.   We provide both distributions for ease of comparison with other predictions  in the literature. Both the source-frame and redshifted $m \to m (1+z)$ masses are given.  The redshifted quantities are the direct effect of the cosmological redshift of the gravitational waves in an expanding universe.  The mass--redshift degeneracy \citep{KrolakSchutz:1987} can be broken by converting the luminosity distance, inferred from the gravitational-wave amplitude, into a redshift, using standard cosmology; this makes it possible to extract source-frame masses \citep[e.g.][]{Haster:2015IMRI}. 

The source frame chirp masses and total masses show practically no dependence on detector sensitivity.  In other words, the distributions of chirp masses and total masses of detectable binaries do not significantly evolve with redshift.   The median source frame chirp masses are  $\mathcal{M}_{c,{\rm full}} = 35^{+10}_{-10}\, M_\odot$ and  $\mathcal{M}_{c, {\rm O1}} =34^{+11}_{-10} \, M_\odot$ for full and O1 detector sensitivity respectively, where the error bars indicate the 90\% confidence intervals.  These values are consistent with the parameters inferred for \TheEvent   $\mathcal{M}_{c,{\rm GW150914}} =  28^{+2}_{-2} \, M_\odot$.  The corresponding total source frame masses are  ${m}_{{\rm tot, full}} = 82^{+21}_{-25} \, M_\odot$ and  ${m}_{{\rm tot, O1}} = 80^{+24}_{-24} \, M_\odot$ respectively, also consistent with the value inferred for \TheEvent   $m_{\rm tot,GW150914} = 65^{+5}_{-4} \, M_\odot$.  \TheEvent is also consistent with the distribution of redshifted chirp masses and total masses, although it resides on the lower side of the redshifted mass distributions, in line with its small inferred redshift. 

In  \autoref{Correlation} we provide two further visualizations of our simulations showing two-dimensional distributions.  In the left-hand panel of  \autoref{Correlation} we show the joint distribution of the mass ratio and redshifted total mass for full design sensitivity and O1 sensitivity.  The inferred ranges for \TheEvent are over plotted.  In the right-hand panels we display the properties of the individual simulated merging binaries in our Monte Carlo simulations, showing the delay time versus chirp mass.  The size and color of the symbols show how much these simulated systems contribute to the overall detection rate. The largest contributions come from binaries in the bottom right of the diagram, i.e., systems with relatively short time delays and high masses which emit stronger gravitational-wave signals, detectable at greater distances.  The preference for short delay times is less strong for the events detectable in the 16 days double-coincident O1 run, which probes a smaller volume, therefore preferring late-time mergers.  The local merger rate instead is dominated by lower mass events with relatively long delay times \citepalias[as can be seen in Figure~9 of][]{Mandel+2016}, and the local detection rate is a trade-off between this and the greater sensitivity to more massive systems.

The predicted distributions show a stronger preference for high masses than either classical population-synthesis predictions for field binary black holes \citep[e.g.,][]{Dominik+2015} or dynamically formed binary black hole models in globular clusters \citep[e.g.,][]{Rodriguez:2015}.  All merging binaries formed through this channel have total masses $\gtrsim 50\Msun$ under the default model assumptions.  Furthermore, we find no delay times shorter than 3\Gyr which has implications for the detectable stochastic background signal \citep{GW150914:stoch}.

\begin{table*}
\caption{Quantification of the impact of model variations on our predictions and a comparison with GW150914.  We list $R_{\rm detect}$, the detection rate at full design sensitivity; $N_{\rm detect}$ (O1), the expected number of detections at the sensitivity of O1 for a 16 day period of double-coincident observations; the median and 90\% intervals for the mass parameters that can be inferred from the waveforms, where $\mathcal{M}_c$ is the chirp mass, $m_{\rm tot}$ the total mass, $q = m_2/m_1$ the mass ratio with component masses $m_1 > m_2$.  For the mass ratio we provide the 90\% lower bound on the $q$. We list the union of the 90\% ranges as ``combined'' parameters.  All parameters refer to the distributions of detectable events at full design sensitivity, unless otherwise indicated.  For comparison, we provide the parameters inferred for GW150914 in the source frame. The reader may also wish to compare with the candidate event mentioned in \citet{GW150914:CBC}, if it is indeed of astrophysical origin.\label{tab1}
}
\begin{center}
\begin{tabular}{ll|cc|cccccl}
\hline 
\hline
 ID & Model &$R_{\rm detect}$(full) 	&$N_{\rm detect}$ (O1)& $\mathcal{M}_c$  & $m_{\rm tot}$ & $q$ &$m_1$       & $m_2$  & Description   \\
 && (yr$^{-1}$)                	& (per 16 days)& $(\Msun)$             & $(\Msun)$                           &        &$(\Msun)$  & $(\Msun)$&\\
\hline
& & &&&&&&&\\
 0    &{\tt DefaultFull} 	& 470 	&-	& $35^{+10}_{-10}$ & $82^{+21}_{-25}$ & $>0.66$ & $44^{+11}_{-15}$ & $36^{+15}_{-10}$ &Standard, full design sensitivity\\[0.15cm]
  0   &{\tt DefaultO1} & -  &1.8 & $34^{+11}_{-10}$ & $80^{+24}_{-24}$ & $>0.68$ & $44^{+12}_{-14}$ & $35^{+15}_{-9}$ &Standard, O1 sensitivity\\[0.15cm]
     
1   &{\tt PoorMixing} 	& 230 	&0.6	& $32^{+10}_{-6}$ & $74^{+24}_{-14}$ & $>0.72$ & $41^{+14}_{-11}$ & $34^{+9}_{-7}$ & Red.~Case M window\\[0.15cm]
2.1&{\tt Zmin0.002} 	& 91 	& 0.3	& $35^{+9}_{-9}$ & $84^{+17}_{-22}$ & $>0.65$ & $47^{+9}_{-14}$ & $35^{+12}_{-9}$ & Red.~metallicity threshold ($0.002$)\\[0.15cm]
2.2&{\tt Zmin0.008} 	& 540 &2.5	& $35^{+9}_{-10}$ & $80^{+20}_{-24}$ & $>0.68$ & $47^{+8}_{-18}$ & $36^{+14}_{-10}$& Inc.~metallicity threshold ($0.008$) \\[0.15cm]
3.1&{\tt ConstA} 	& 1200 &1.4	& $34^{+10}_{-11}$ & $79^{+22}_{-25}$ & $>0.68$ & $42^{+14}_{-14}$ & $35^{+13}_{-10}$  &Slow winds (fixed sep.)  \\[0.15cm]
3.2&{\tt HalvedA} 	& 1000  &1.2	& $34^{+10}_{-11}$ & $78^{+23}_{-25}$ & $>0.69$ & $44^{+10}_{-16}$ & $35^{+12}_{-10}$ &Slow winds (halving sep.)   \\[0.15cm]
4.1&{\tt Mdot2} 		& 0.0 	& 0.0		& - & - & - & - & - &Enh.~mass loss (doubled) \\[0.15cm]
4.2&{\tt Mdot2ConstA} & 620  &	1.5	& $26^{+14}_{-12}$ & $59^{+32}_{-27}$ & $>0.55$ & $34^{+15}_{-17}$ & $26^{+19}_{-11}$ & Enh.~mass loss \& slow winds \\[0.15cm]
4.3&{\tt Mdot0.2} 	& 1500 &1.6  	& $39^{+11}_{-9}$ & $91^{+23}_{-22}$ & $>0.74$ & $50^{+10}_{-14}$ & $42^{+14}_{-9}$ & Red.~mass loss (by factor of 5)   \\[0.15cm]
5   &{\tt PISN80} 	& 600 	&2.1 	& $40^{+8}_{-16}$ & $93^{+17}_{-37}$ & $>0.59$ & $51^{+16}_{-19}$ & $37^{+18}_{-11}$ & Enh.~PISN threshold ($80\Msun$) \\[0.15cm]
6   &{\tt Dex0.5} 	& 1400 & 10	& $34^{+10}_{-10}$ & $77^{+24}_{-22}$ & $>0.71$ & $43^{+11}_{-14}$ & $37^{+13}_{-11}$ & Enh.~metallicity spread (0.5 dex) \\[0.15cm]

\hline 

\multicolumn{2}{l|}{Combined}&  0--1500& 0--10& 14--50 & 32--114 & >0.55 & 17--67 & 15--56 & Union of 90\% ranges\\[0.15cm]
\hline 

\multicolumn{2}{l|}{GW150914} 
  & 
& 1 &
$28^{+2}_{-2}$ &
$65^{+5}_{-4}$ &
$>0.65$ &
$36^{+5}_{-4}$ & 
$29^{+4}_{-4}$& \citet{LIGO+2016_properties}\\[0.15cm]

\hline
\end{tabular}
\end{center}
\end{table*}

\section{Robustness of results}\label{variations}

Substantial uncertainties in these simulations arise from several sources: (i) the assumptions for the initial conditions, (ii) the physics of the evolution of the systems (in particular the efficiency of the mixing processes, the mass and angular momentum losses), (iii) cosmological assumptions, and (iv) assumptions regarding gravitational-wave detectability.   The impact of the initial conditions such as the binary fraction and the adopted distribution functions for the binary parameters has been quantified by \citet{de-Mink+2015} for the classical isolated binary scenario.  They concluded that the impact of uncertainties in the initial distributions is fully dominated by uncertainties in the initial mass function, which accounts for a factor of 8 up and down in the overall rate, with very little to no effect on the distribution of the properties of double black hole mergers. 
In \citetalias{Mandel+2016}  we quantified various aspects of the impact of (ii) and (iii) on the local ($z=0$) merger rate, as well as 
the maximum cosmological merger rate and the redshift at which the maximum is reached.  Here, we provide a similar exploration, now probing the impact of model variations on the detection rate for full design sensitivity observations $R_{\rm detect}$(full), the number of detections expected for the 16 days of double-coincident O1 observations analyzed so far $N_{\rm detect}$ (O1) \citep{GW150914:CBC}, as well as the median and 90\% confidence intervals on the chirp mass, total mass, mass ratio and component masses in the source frame.  

\subsection{Model variations}
\rr{We consider the same variations as \citetalias{Mandel+2016}, to which we refer for an extensive discussion and motivation for the considered variations. Here, we limit ourselves to a brief summary. Results for all model variations are summarized in \autoref{tab1}.}  

\rr{The efficiency of the mixing processes constitutes one of the main uncertainties in the evolutionary models. We therefore consider a variation  {\tt PoorMixing} in which we used a more conservative threshold for chemically homogeneous evolution that roughly halves the window of interest in initial orbital period space. We vary the threshold metallicity for chemically homogeneous evolution in models {\tt  Zmin0.002}  and  {\tt Zmin0.008}.  We consider the uncertainties in angular momentum loss driven by stellar winds in models  {\tt ConstA}, which represents enhanced angular momentum loss by keeping the separation fixed, and {\tt HalvedA}, which further enhances angular momentum loss under the assumption of slow winds, shrinking the orbital separation by a factor of two. Model variations {\tt Mdot2} and {\tt Mdot0.2} represent enhanced and reduced mass loss. These variations account for uncertainties in the wind mass loss as well as other modes of mass loss such as eruptive mass loss episodes expected for pulsational pair instability supernovae.  Model {\tt Mdot2ConstA} considers enhanced mass loss but assumes higher angular momentum loss from the system \citepalias[see sect.~7.2 and 7.3 of][]{Mandel+2016}. Finally we consider the uncertainty in the threshold for pair instability supernova in model {\tt PISN80} and a variation in the assumed metallicity spread at each redshift in model {\tt Dex0.5}. }

One variation, model {\tt Mdot2}, which corresponds to a doubling of the mass loss, predicts no detectable events. \rr{We do expect that detectable events would arise even with doubled mass loss at lower metallicities, given the $\propto Z^{0.85}$ scaling of wind-driven mass loss rates \citep{Vink+2001}, consistent with the findings of \citet{Marchant:2016} who analyzed $Z=Z_\odot /10,\, Z_\odot /20,\, Z_\odot /50$ populations.  Our present results, which are based on $Z=0.004$ models, represent a very conservative assumption.} 

The predictions for the detection rate at full sensitivity vary substantially, $R_{\rm detect}$(full) $= 90-1500$ per year for all models with non-zero predictions. The prediction for the number of detections in the 16 days of double-coincident O1 data analyzed so far varies between $N_{\rm detect}$(O1) $= 0.3-2.5$  with only two exceptions:  model {\tt Mdot2} predicts no detections and model {\tt Dex0.5} predicts 10 detections. 

We find that the preference for relatively high chirp and total masses is a robust prediction seen in all model variations that yield detectable events.  Model variation {\tt Mdot2ConstA} in which we adopted enhanced wind mass loss results in the lowest median masses.  Model  {\tt PISN80} results in the highest median masses. This model increases the maximum final mass at which stars can  become black holes (instead of exploding in a pair instability supernova which would leave no remnant).  

The preference for comparable masses is also a robust prediction.  The preference for equal masses becomes stronger when we consider a reduced efficiency of tidally induced mixing {\tt PoorMixing},  $q>0.72$, and it is least strong for the model with reduced mass loss and enhanced angular momentum loss through slow winds {\tt Mdot2ConstA}, $q>0.55$.

At this time, all models apart from {\tt Dex0.5} are consistent with the number of detections observed during the first 16 days of double-coincident observation from the O1 run after accounting for Poison statistics and the possibility that this channel is not the only channel that contributes to binary black hole detections.

\subsection{Uncertainties in the gravitational-wave detectability}\label{GWdetect}

We use a single-detector signal-to-noise ratio threshold of $8$ as a proxy for detectability by the LIGO-Virgo network.  In practice, gravitational-wave search pipelines \citep[e.g.,][]{ihope,Cannon:2012} use more complex statistics than the signal-to-noise ratio to treat non-stationary, non-Gaussian noise backgrounds.  As a result, the actual sensitivity of advanced gravitational-wave detectors will depend on the details of the network (such as the number of detectors operating in coincidence at a given time, which in turn depends on their duty factor), the detector data quality, the specific algorithms used for the search, and even the details of the source, such as the component masses.  

Moreover, the spins of the binary components can have a significant effect on the gravitational-wave signal.  This is particularly true for massive binaries: large aligned spins can enhance the strength of the signal, possibly increasing the detector sensitive volume by factors of $\sim 2$ \citep[see, e.g., figure 6 of][]{Belczynski:2014VMS}.  Therefore, our detectability predictions are simplifications which must be treated with caution; however, the uncertainties involved are likely smaller than those in the physics governing the evolution of the binary systems. 

The detection rate predictions are based on advanced LIGO detectors operating at either full sensitivity or O1 sensitivity.  The detectors will gradually evolve in sensitivity between 2015 and the end of the decade, with several scheduled data-taking runs interspersed with commissioning breaks \citep{scenarios}.  While the exact predictions for any intermediate runs depend on the exact shape of the detector noise spectrum and must take into account the cosmological variations in merger rates as described above, a crude estimate can be made by assuming that the detection rate scales with the surveyed volume \citep[see Fig.~4 of][]{LIGO+2016_astroImplications}.

\section{Summary and Conclusion}

{The channel for chemically homogeneous evolution in tidally distorted massive binary systems is of large interest in light of current searches for binary black hole mergers. The high component masses and comparable masses for the components inferred for GW150914 are a natural and robust outcome of this evolutionary channel. The predicted detection rate is less certain but fully consistent with the first 16 days of double-coincident O1 observations.} 

{ At present, with a single confident detection, it is not possible to distinguish between this channel, the classical channel for isolated binary evolution, and the dynamical formation channel. However, the near future prospect of up to hundreds of detections per year \citep{LIGO+2016_rate} will probe the demographics of stellar mass binary black holes. This, together with measurements of the stochastic background from individually unresolvable mergers \citep[e.g.][]{GW150914:stoch}, will provide constraints on the formation mechanisms. }  

{ Our default model predicts about 500 detections per year at full design sensitivity, corresponding to about 1.8 detections in 16 days of O1-sensitivity data.  The model variations we consider give variations by factors of 3--5 up or down, although the possibility of zero detections from this channel can not be excluded at this stage.} 
  
{ The preference for binary black hole mergers with comparable component masses is a robust outcome of all models considered here (in all model variations we find that 90\% of the detectable events have mass ratios $q$ larger than $0.55$).  The same holds for the preference for high total and chirp masses (the median total mass ranges from $59$ to $93\, \Msun$ in the model variations we consider).}
  
{Possible future detections of binary black hole mergers with significantly unequal component masses or low total masses will be evidence in favor of contributions by the classical isolated binary channel and/or the dynamical formation channel.  At 90\% confidence, none of the model variations predict total masses below $32\, M_\odot$ or mass ratios $q<0.55$.  Observations outside these boundaries are unlikely to arise from this channel in which both stars evolve chemically homogeneously. However, it would be interesting to explore variations of this evolutionary path in which only one of the stars evolves chemically homogeneously.} 
  
{ We find that the cosmological merger rate peaks at a redshift of 0.4 with the majority of events being just out of reach of the full design sensitivity of the detectors. We find no mergers beyond $z=1.5$ in the default model. This has implications for the stochastic background signal that can be tested against predictions from other binary black hole formation channels.}
  
{ Although the simulated merger and detection rates for this channel are sensitive to model uncertainties, this channel does not suffer from the key physics uncertainties that affect the classical isolated binary evolutionary channel, namely the treatment of unstable and non conservative mass transfer, common envelope ejection events, and the still unconstrained black hole birth kicks. Further efforts are needed to advance detailed one-dimensional \citep[such as][]{Marchant:2016} and three-dimensional simulations of the physical processes affecting massive near-contact binaries. If future disentangling of the contributions by different scenarios becomes possible, gravitational-wave events will provide interesting constraints on the unique physics of the mixing processes that govern this channel. }

\acknowledgements
\vspace*{0.5cm}
Both authors contributed equally to this work.  The authors are grateful to Yuri Levin, Colin Norman \rr{and the anonymous referee} for comments on the manuscript. The authors further acknowledge the Leiden Lorentz Center workshop ``The Impact of Massive Binaries Throughout the Universe'' and the Netherlands Research School for Astronomy (NOVA) for a visitor grant for IM. SdM acknowledges support by a Marie Sklodowska-Curie Action Incoming Fellowship (H2020 MSCA-IF-2014, project id 661502). 
\bibliography{my_bib,Mandel}

\end{document}